\documentclass[runningheads]{llncs}

\usepackage{cite}
\usepackage{comment}
\usepackage{amsmath,amssymb,amsfonts}
\usepackage{algorithmic}
\usepackage{graphicx}
\usepackage{textcomp}
\usepackage[svgnames]{xcolor}
\def\BibTeX{{\rm B\kern-.05em{\sc i\kern-.025em b}\kern-.08em
    T\kern-.1667em\lower.7ex\hbox{E}\kern-.125emX}}
\usepackage{subfigure}
\usepackage{geometry}
\usepackage{indentfirst}

\usepackage{hyperref}
\usepackage{csquotes}

\usepackage[inline]{enumitem}
\newlist{inlineenum}{enumerate*}{1}
\setlist[inlineenum]{label=(\arabic*),
 itemjoin={{; }},
}

\usepackage{hyperref}
\hypersetup{
    colorlinks=true,
    linkcolor=blue,
    citecolor=blue,
}

\usepackage{wrapfig}
\usepackage{seqsplit}

\begin{document}

\author{
    Benjamin Michalowicz\inst{1} \and Eric Raut\inst{1} \and
    Yan Kang\inst{1} \and Tony Curtis\inst{1} \and  Barbara Chapman \inst{1,2} \and Dossay Oryspayev\inst{2}}
    \institute{Institute For Advanced Computational Science, Stony Brook University, NY, USA 
    \email{\{benjamin.michalowicz, eric.raut, yan.kang, anthony.curtis, barbara.chapman\}@stonybrook.edu}
    \and
    Computational Science Initiative, Brookhaven National Laboratory, NY, USA \\
    \email{\{bchapman, doryspaye\}@bnl.gov}}

\authorrunning{B.T.~Michalowicz, E.~Raut, Y.~Kang et al.}

% \title{On experiences with, and the behavior of, the OpenMP
% implementations on Ookami and comparison to that of Fugaku*\\
% \thanks{Identify applicable funding agency here. If none, delete this.}
% }

\title{Comparing OpenMP Implementations With Applications Across A64FX Platforms}

\maketitle
\begin{abstract}
The development of the A64FX processor by Fujitsu has created a massive innovation in High-Performance Computing and the birth of Fugaku: the current world's fastest supercomputer. A variety of tools are used to analyze the run-times and performances of several applications, and in particular, how these applications scale on the A64FX processor. We examine the performance and behavior of applications through OpenMP scaling and how their performance differs across different compilers both on the new Ookami cluster at Stony Brook University as well as the Fugaku supercomputer at RIKEN in Japan.
\end{abstract}

% \begin{IEEEkeywords}
% HPC, OpenMP, Parallel Programming, Supercomputing, A64FX
% \end{IEEEkeywords}

\section{Introduction}
The introduction of the A64FX processor by Fujitsu, and its use in the Fugaku supercomputer (Fugaku), has sparked the re-emergence of vectorized processors/programming and the birth of the next world's-fastest supercomputer~\footnote{\url{https://top500.org/}}. This comes on top of the fact that the A64FX chip also brings an unprecedented co-design approach, impressive performance, and energy awareness that puts it at the top of all 5 major HPC benchmarks. In this paper, we will be analyzing OpenMP~\footnote{\url{https://www.openmp.org/}}, a well-known shared memory/parallel programming model, from its scaling abilities on the A64FX processor to how it performs across different compiler toolchains.

The full list of current compilers that support OpenMP can be found at \texttt{https://openmp.org}\footnote{\url{https://www.openmp.org/resources/openmp-compilers-tools/}}.
Although there is one OpenMP specification, compiler support varies both in terms of specific OpenMP features and general performance.

In the next two subsections we give a brief overview of the A64FX processor, followed by the paper's contribution and organization.

% \subsection{OpenMP Overview}
% OpenMP is a directive-based standard for parallel programming on shared memory systems. The ease of use and, comparatively, less code refactoring while utilizing it on shared memory systems makes OpenMP very attractive for obtaining efficient parallel versions of serial programs. The programmer can use compiler directives, library routines, and environment variables
% % TONY: which are specified in OpenMP specification,
% to write parallel programs for shared memory systems in languages like Fortran and C/C++. The OpenMP ARB (Architecture Review Board) is composed of diverse members from different companies, and governs and actively works on promoting the OpenMP API in supercomputing and other communities.

%\subsubsection{Compiler vs. Runtime}

\subsection{The A64FX Processor}
The A64FX processor~\cite{RyohiOkazaki2020,Sato2020} is the processor specifically manufactured for Fugaku, which was made possible as part of the Japanese FLAGSHIP 2020 project as a co-design between RIKEN and Fujitsu. Currently, Fugaku is ranked number 1 on both Top500 and HPCG lists. The A64FX, is a general-purpose processor based on the Armv8.2-A architecture~\cite{Sato2020} and comes with 48 compute cores + 2/4 cores dedicated to OS activities.

The A64FX processor produced by Fujitsu
% TONY: is comprised of
has
4 core memory groups (CMG). In the FX700 chip, each CMG has 12 cores, while the FX1000 chip has 2-4 extra assistant cores. Ookami currently has the FX700 chips, with each core laid out sequentially: cores 0-11 make up CMG 0, 12-23 make up CMG 1, etc.~\cite{FujitsuMan}. 

\subsection{Paper's contribution and organization}

Although OpenMP support is available in many compilers, to the best of our knowledge, there were no studies of OpenMP's performance in various compilers (and specific versions) specifically for A64FX processors with the set of applications considered in this paper, and features of OpenMP they're using. To that end, the paper's contributions are as follows:
\begin{itemize}
    \item We present and evaluate the single node performance of various applications using all available compilers on two systems that have A64FX processors, viz. Ookami and Fugaku.
    \item We present, evaluate, and compare the differences in performance on two different models of A64FX.
    \item We discuss our findings, and based on the results obtained, we summarize the maturity level of compilers available on these two systems to fully utilize the features of A64FX processors.
\end{itemize}

The rest of the paper is organized as follows. In section~\ref{sect:list_of_applications_and_setup} we present the details of the applications considered in this study, and of the systems and compilers used. In section~\ref{sect:experimental_results} we present and discuss the results obtained as well as inferences obtained from running applications through various profilers and performance analysis tools. In section~\ref{sect:related_work} we list related work and discuss their contributions and the contribution of our work. Finally, in section~\ref{sect:conclusions_and_future_work} we summarize our findings and list some work to be performed in near future.

\section{List of applications and experimental setup} \label{sect:list_of_applications_and_setup}

\subsection{List of Applications}

\begin{itemize}
    \item PENNANT~\cite{LANLPennant} - A mesh physics mini-app designed for advanced architecture research. PENNANT is dominated by pointer chasing and operates based on input files with different parameters. The larger the parameters, the larger the mesh. PENNANT can be run solely with MPI, OpenMP, or in a hybrid MPI+OpenMP fashion, and uses OpenMP's static scheduling feature. 
    \item SWIM -  a weather forecasting model designed for testing current performance of supercomputers. It is a Fortran code using OpenMP. Like PENNANT, it also uses static scheduling of OpenMP. It has been updated within SPEC CPU 2000 benchmark collections by Paul N. Swarztrauber~\cite{Swim}.
    \item Minimod~\cite{Meng2020,Raut2020,raut2021porting} - a seismic modeling mini-app that solves the acoustic wave equation using finite differences with a stencil. Minimod is developed by TotalEnergies and is designed as a platform to study the performance of emerging compilers and runtimes for HPC. In this paper we consider the OpenMP loop-based and task-based variants of the code~\cite{Raut2020}.
    
\end{itemize}

\subsection{Systems and Compilers}\label{SystemCompiler}

\subsubsection*{Fugaku} is the world's fastest supercomputer, located at the RIKEN Center for Computational Science in Japan~\cite{R-CCS}, and runs on the FX1000 A64FX, which provides extra cores for OS-communication. Its underlying TofuD interconnect is implemented as an interconnect controller (ICC) chip to allow for low latency and offloading.\cite{icc} 
%{R}{0.4\linewidth}     

\begin{wraptable}{r}{8.5cm}
%\begin{table}[h!]
\centering
\vspace{-.75\intextsep}
\hspace*{-.75\columnsep}
  %\begin{center}
    \begin{tabular}{|c|c|c|} \hline
      \multicolumn{1}{|c}{} & \multicolumn{2}{|c|}{\textbf{Versions}}\\
      \hline
      \textbf{Compiler Family} & \textbf{Fugaku} & \textbf{Ookami} \\
      \hline
      ARM & - & $20.3$\\
      Cray & - & $10.0.1$\\ % <-- added row here
      Fujitsu & $4.3.0\text{a}$, $4.4.0\text{a}$ & -\\
      GCC & $8.3.1$, $10.2.1$ & $8.3.1$, $10.2.1$, $11.0.0$\\
      LLVM & $11.0.0$ & $11.0.0$, $12.0.0$\\
      \hline
    \end{tabular}
    \caption{Compilers of Fugaku and Ookami.}
    \label{tab:table1}
  %\end{center}
%\end{table}
\vspace{-1.25\intextsep}
\hspace*{-.75\columnsep}
\end{wraptable}

\subsubsection*{Ookami} is a cluster installed at Stony Brook University (SBU) in the middle of 2020. It contains 174 compute nodes, with another two set aside for quick experimentation. Ookami was funded through an NSF grant~\cite{NSF-1927880} as the first A64FX cluster outside of Japan. It comes with an array of software modules, including GNU, LLVM, and Cray compilers, profilers, and MVAPICH/OpenMPI packages. Ookami uses a non-blocking HDR 200 switching fabric via 9 40-port Mellanox Infiniband switches in a 2-level tree, which allows for a peak bandwidth of 100 Gb/s between nodes. In addition, each node currently has 32GB of high-bandwidth memory with a peak memory bandwidth of 1 TB/s. Both systems' compiler toolchains are shown in Table~\ref{tab:table1}.

%\begin{center}
%    \begin{tabular}{ |c|c|c|c|c|c }
%    \centering
%        \hline
%        \textbf{Compiler Family (Ookami)} & \textbf{Versions} & \textbf{Compiler Family (Fugaku)} & \textbf{versions} \\
%        \hline
%        Cray & 10.0.1 & Fujitsu & 4.3.0a\\
%        GCC & 8.3.1, 10.2.1, 11.0.0 & GCC & 8.3.1,10.2.1\\
%        LLVM & 11.0.0, 12.0.0  & LLVM & 11.0.0\\
%        ARM & 20.3 & & \\
%        \hline
%    \end{tabular}
%    \caption{Compiler Toolchain of two systems.}
%    %\label{tbl:compiler_toolchain_of_systems}
%\end{center}

\subsection{Runtime Environment}

Each benchmark was run on 1 compute node with 1 MPI rank/process to avoid shared memory operations that occur with 2 or more processing elements and over-subscription of threads to cores, which result in degraded performance.
Threads are bound to cores using the \texttt{\seqsplit{OMP\_PLACES}} environment variable.
%assigned to \texttt{\seqsplit{"{start\_core}:num\_cores"}}.

Threads are assigned to specific cores (e.g. Thread 0 is assigned to Core 0) and divided equally among specific CMGs. For example, 32 threads are divided equally among the four CMGs on a single Ookami node (cores 0-8 in CMG 0, 12-19 in CMG 1, etc.) using \\
\indent \texttt{OMP\_PLACES="\{0\}:8,\{12\}:8,\{24\}:8,\{36\}:8"}. \\
We ran experiments using 1, 2, 4, 8, 12, 16, 24, 32, 36, and 48 OpenMP threads. For every value up to 12, we placed all threads in one CMG. The 16-thread and 24-thread experiments were run on 2 CMGs, with each group having half the total thread values. The 32-thread and 48-thread experiments were run on all 4 CMGs on the A64FX chip, with 36 threads being run on 3 CMGs.

\begin{wraptable}{r}{8.5cm}
%\begin{table}[h!]
%\begin{center}
\centering
\vspace{-\intextsep}
\hspace*{-.75\columnsep}

    \begin{tabular}{ |c|c|l| }
        \hline
        \textbf{Compiler} & \textbf{Flags} \\
        \hline
        Cray & \parbox[t]{5cm}{\texttt{-homp -hvector3 -hthread3}} \\
        \hline
        GCC & \parbox[t]{5cm}{\texttt{-mcpu=a64fx \\-Ofast -fopenmp}}\\ 
        \hline
        LLVM & \parbox[t]{5cm}{\texttt{-mcpu=a64fx \\-Ofast -fopenmp}} \\ 
        \hline 
        Fujitsu-Traditional & \parbox[t]{5cm}{\texttt{-Nnoclang -Nlibomp -O3 -Kfast,\\-Kopenmp,ARMV8\_2\_A\\-KSVE,A64FX}} \\ 
        \hline
        Fujitsu-LLVM & \parbox[t]{5cm}{\texttt{-Nclang -Nlibomp -Ofast -Kfast,openmp -mcpu=a64fx+sve}}\\
        \hline
    \end{tabular}
    % \caption{Set of explicitly set flags for each of the compilers.}
    \caption{Flags used for each compiler.}
    \label{tab:table2}
%\end{center}
%\end{table}
\vspace{-1.9\intextsep}
%\hspace*{-.75\columnsep}

\end{wraptable}

\subsection{Compiler options}\label{Optimizations}
For each compiler mentioned in
Section~\ref{SystemCompiler}, we turned on specific flags, maximizing thread optimization, SVE instruction generation, and execution speed while maintaining correctness of output. We also enabled fine-tuning for the A64FX processor and the ARM-8.2 architectures where possible. The flags are listed for each compiler/group are set as shown
%\footnote{For any GNU and LLVM compiler: If compiling directly on an A64FX node, use \texttt{-mcpu=native} instead.}
in Table~\ref{tab:table2}. Note that GCC versions before version 9 do not support the \texttt{mcpu=a64fx} flag -- for GCC 8, we compile directly on an A64FX node and use \texttt{mcpu=native}. These flags instruct the compiler to use auto-vectorization; we have not tested OpenMP's SIMD clauses.

\section{Experimental results} \label{sect:experimental_results}
Our experiments analyzed runtime, relative speedup through OpenMP threads, and efficiency with respect to different compiler/compiler classes -- Cray, ARM, GNU, and LLVM. Subsection~\ref{OokamiRes} contains all the results run on SBU's Ookami cluster, followed by subsection \ref{FugakuRes} containing results from Fugaku. For each application, we deemed three compilers as "best in class" (best runtime) for the families mentioned above: GNU-10.2.0 (\texttt{\seqsplit{gcc/g++/gfortran}}), \texttt{\seqsplit{ARM/LLVM-20.1.3}} (\texttt{\seqsplit{armclang/armclang++/armflang}}), and \texttt{\seqsplit{Cray-10.0.1}} (\texttt{\seqsplit{cc/CC/ftn}}), with the results for the other compilers explained in the following subsections. Preliminary results are showin in \cite{Michalowicz21}.

Our results are drawn from running our programs 5 times per OpenMP thread value requested (1, 2, 4, 8, 12, 16, 24, 32, 36, 48) and taking the arithmetic mean values from each set of runs. These experiments are limited to a single node.

\subsection{Ookami}{\label{OokamiRes}}

\subsubsection{PENNANT} Runs were based on 2 medium-sized inputs whose memory constraints did not expend the A64FX's high bandwidth memory and swap space: \textbf{Leblancbig} and \textbf{Sedovbig}. These inputs both revolve around structured meshes with all square zones, but deal with considerably different mesh parameters, such as the number of elements in the respective mesh's zone adjacency lists and the number master/slave points/array sizes.

\begin{figure}[!htb]

  \centering
    \subfigure[Compiler Runtime Comparisons]
    {\includegraphics[width=0.411\textwidth]{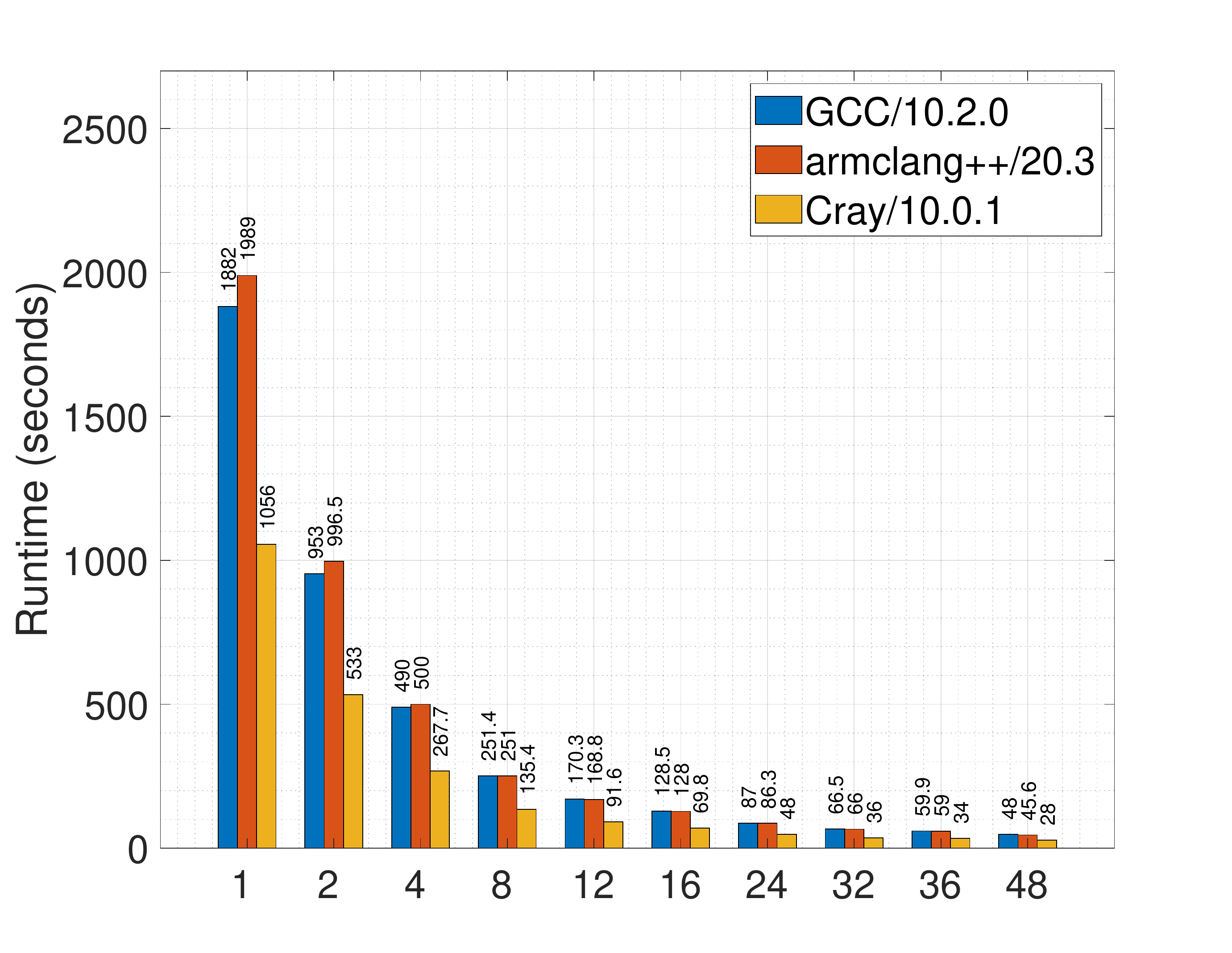} \label{fig:leblancbig-compilers}}
    \subfigure[Relative Speedup]{\includegraphics[width=0.411\textwidth]{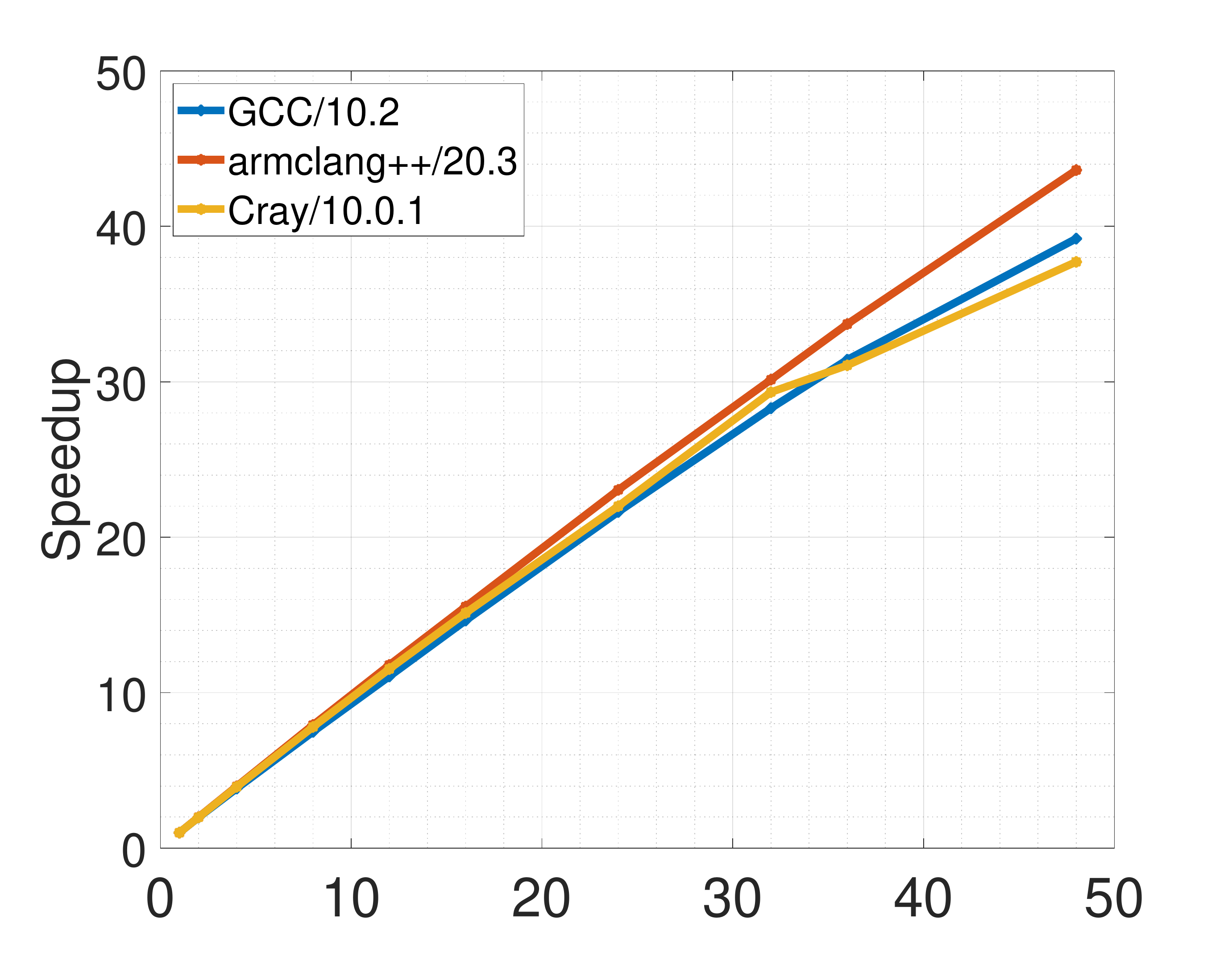} \label{fig:leblancbig-speedup}}
    \caption{PENNANT/LeblancBig Input Results on Ookami; X-axes refer to number of OpenMP threads}
\end{figure}

Figure \ref{fig:leblancbig-compilers} shows how, in every value given for \texttt{OMP\_NUM\_THREADS}, the Cray compilers vastly outpace every other compiler toolchain presented. It has a maximum runtime with \textbf{Leblancbig} of 1056 seconds on 1 OpenMP thread, and 28 seconds on 48 threads. Conversely, the generic LLVM compilers had the absolute worst runtime, running consistently around 2200 seconds on 1 OpenMP thread, nearing 70 seconds with 48 threads. Part of this is a result of how many SVE instructions are generated by each compiler when vector optimizations are turned on at the compilation stage. Cray generates the second largest amount of SVE instructions after the Fujitsu compilers, followed by the generic GNU and ARM compilers. Conversely, the generic LLVM compilers produce no SVE instructions at all, nor do they make use of the A64FX's \texttt{z[0-32]} registers.

In Figure \ref{fig:leblancbig-speedup}, we show the relative speedup observed between each of the \enquote{best in class compilers}, measured by the amount of speedup at a given thread value compared to 1 OpenMP thread. While the \texttt{armclang} results from Figure \ref{fig:leblancbig-compilers} show it having the slowest runtime, it has the largest and most linear relative speedup, with the Cray compilers having the smallest relative speedup. Because the Cray compiler is able to efficiently utilize SVE instructions, increasing the number of OpenMP threads will not necessarily guarantee a linear speedup.

Efficiency for these tests is measured as speedup divided by the number of threads used for a given result. We noticed that that the ARM compilers are the most efficient when compared to the GNU and Cray compilers.

\begin{figure}[!htb]

  \centering
    \subfigure[Compiler Runtime Comparison]{\includegraphics[width=0.411\textwidth]{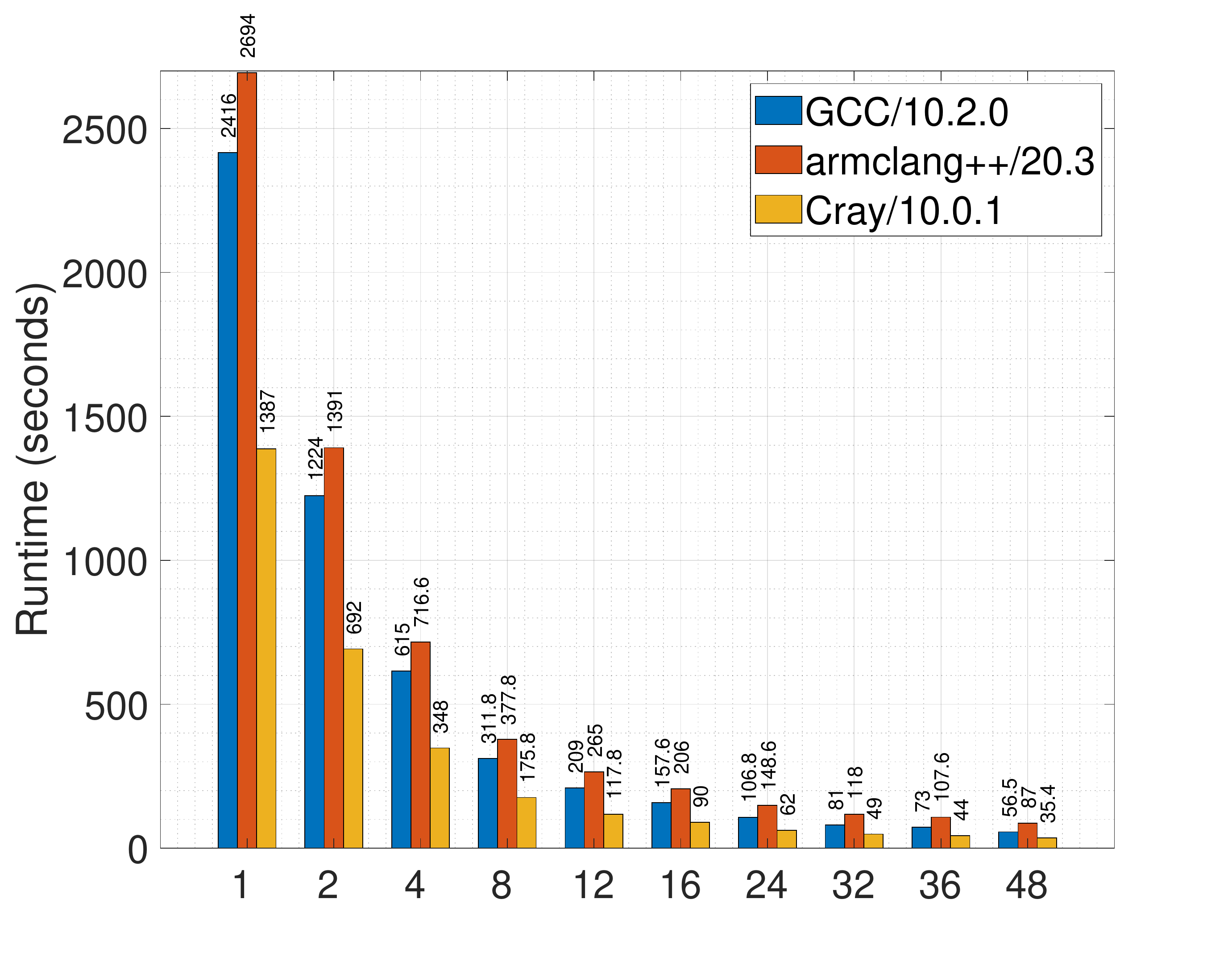} \label{fig:sedovbig-compilers}}
    \subfigure[Relative Speedup]{\includegraphics[width=0.411\textwidth]{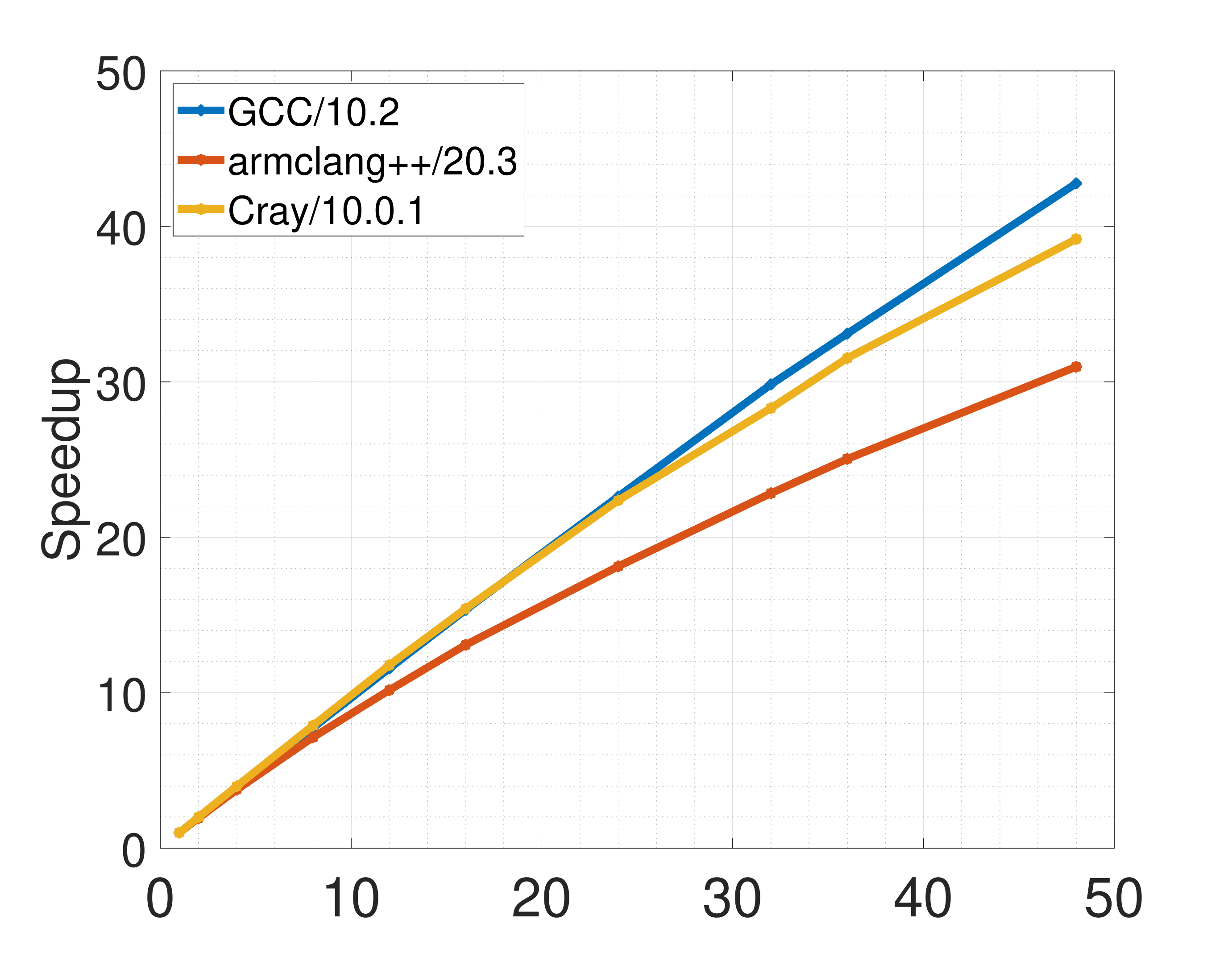} \label{fig:sedovbig-speedup}}
    \caption{PENNANT/SedovBig Input Results on Ookami; X-axes refer to number of OpenMP threads}
\end{figure}

Similar trends are seen with \textbf{Sedovbig}, with the fastest runtimes seen with the Cray compilers--its slowest runtime being 1387 seconds--and the slowest runtimes seen by the ARM-based LLVM compilers (2694 seconds)--in Figure \ref{fig:sedovbig-compilers}. The generic LLVM compilers, not shown in the graph, displayed runtimes as slow as 3000 seconds.

Unlike the results from Figure \ref{fig:leblancbig-speedup}, Figure \ref{fig:sedovbig-speedup} shows the GNU compilers having the most linear/largest speedup, with the Cray compilers coming in relatively close until diverging at the 24-thread experiments. The armclang experiments deviate from the Cray and GNU experiments after 4 OpenMP threads.

\subsubsection{SWIM} was run with the default test problem, \texttt{swim.ref.in}. It sets up a 7701x7701 matrix running 3000 iterations. In our experiments, we tested 7 different compiler versions, but to avoid clutter and data overlap, we have chosen 3 representatives from the various compiler families:  GNU, ARM's LLVM-based compiler, and Cray.  We present results from these compilers in this section. The runtime results and speed-up plots are shown in Figures~\ref{fig:swim-3} and \ref{fig:swim-speedup}:

\begin{figure}[!htb]

%%Below: Idea for multiple graphs on one page, each one could have its own label
  \centering
    \subfigure[]{\includegraphics[width=0.411\textwidth]{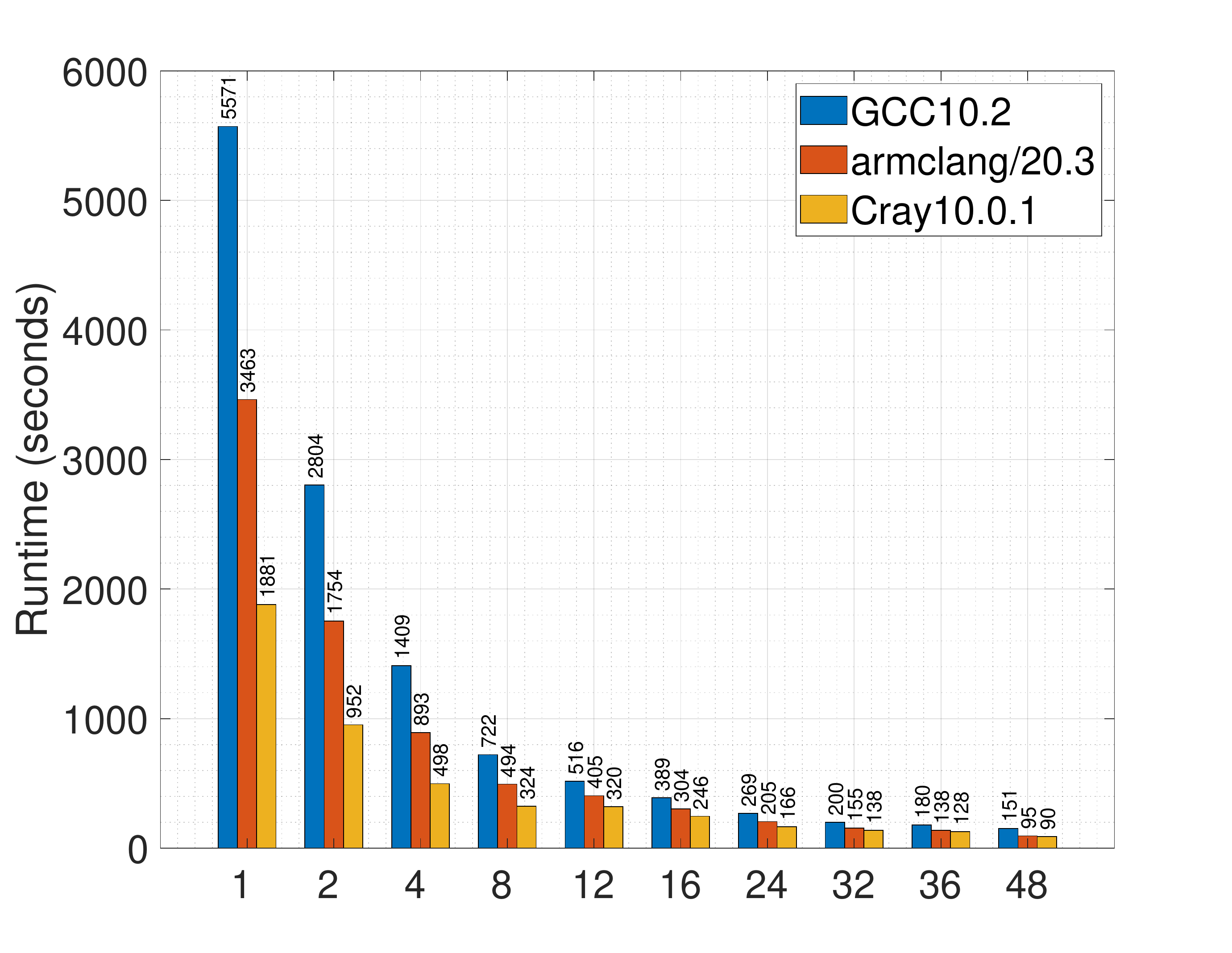} \label{fig:swim-3}}
    \subfigure[]{\includegraphics[width=0.411\textwidth]{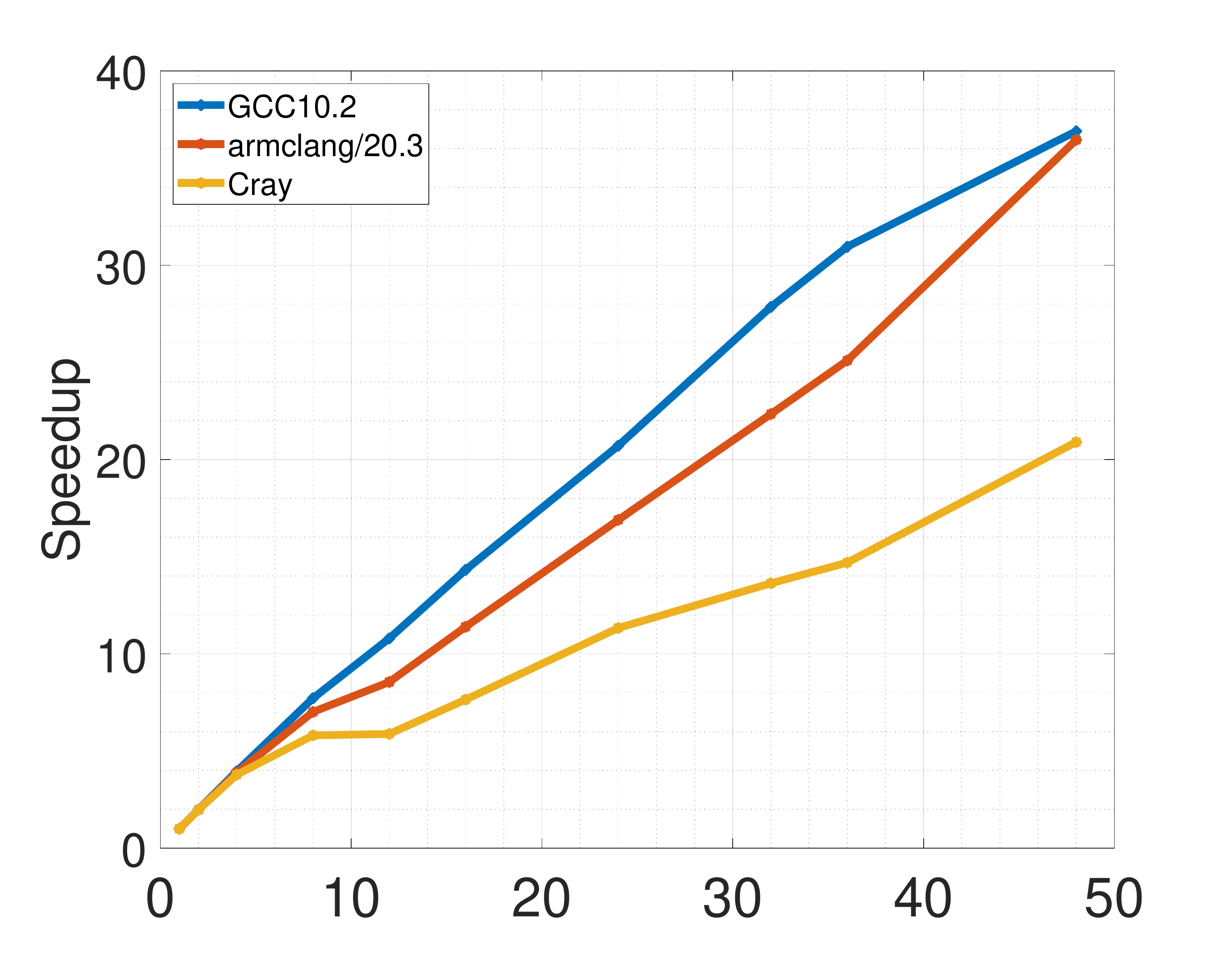} \label{fig:swim-speedup}}
    \caption{(a) SWIM/Ookami Compiler Runtimes (b) SWIM/Ookami Relative Speedup}

\end{figure}

As shown in Figure~\ref{fig:swim-3}, it is clear that the Cray compiler has the best performance among all three compilers with SVE support enabled on the Ookami cluster. Cray obtained a 2.3x faster performance than \texttt{armclang} with a single thread. The ARM-based LLVM compiler generally has better performance than the ARM-based GNU compilers, which is also expected. The LLVM compiler has generated more SVE operations comparing with GNU compiler which leads to the better runtime performance. 

In Figure~\ref{fig:swim-speedup}, we see that among all the compilers, the GNU compiler seems to have the greatest speed up. 48 threads achieve a 37x speed-up over 1 thread. Conversely, the Cray compiler only gives a 16x speedup between 1 and 48 threads shown in Figure~\ref{fig:swim-speedup}. 

A clear result shows that the GNU compiler obtained the best efficiency among all these three compilers. On the contrary, the Cray compiler seems to have a much lower efficiency than GNU and ARM based compilers for all runs. For all three compilers the general trends are the similar. Clear drops are happening when using 2, 4, 8 and 12 OpenMP threads especially at 12 threads. It is also interesting that even though Cray compiler obtained the best runtime performance, there are still a lot more could be improved.

With profiling tools ARM MAP and CrayPat \cite{CrayPat} on Ookami with 48 threads, SWIM has spent 70.2\% runtime on OpenMP region which is understandable since it is a purely OpenMP benchmark. OpenMP generates a small amount of overhead: 28.5\% was seen with this particular runs. Like with PENNANT, this indicates that more threads will spend more time communicating with each other and less overall time on performing computations.

\subsubsection{Minimod} runs with the following two different OpenMP configurations (see \cite{Raut2020} for details):
\begin{itemize}
    \item Loop xy: Grid is blocked in $x$ (largest-stride) and $y$ dimensions. A OpenMP \texttt{parallel for} loop is applied to the 2-D loop nest over x-y blocks. (A \texttt{collapse(2)} is used to combine the two loops).
    \item Tasks xy: Grid is blocked in $x$ and $y$ dimensions. Each x-y block is a task using OpenMP's \textit{task} directive. OpenMP's \texttt{depend} clause is used to manage dependencies between timesteps.
\end{itemize}

\begin{figure}[!htb]
  \centering
    \subfigure[]{\includegraphics[width=0.411\textwidth]{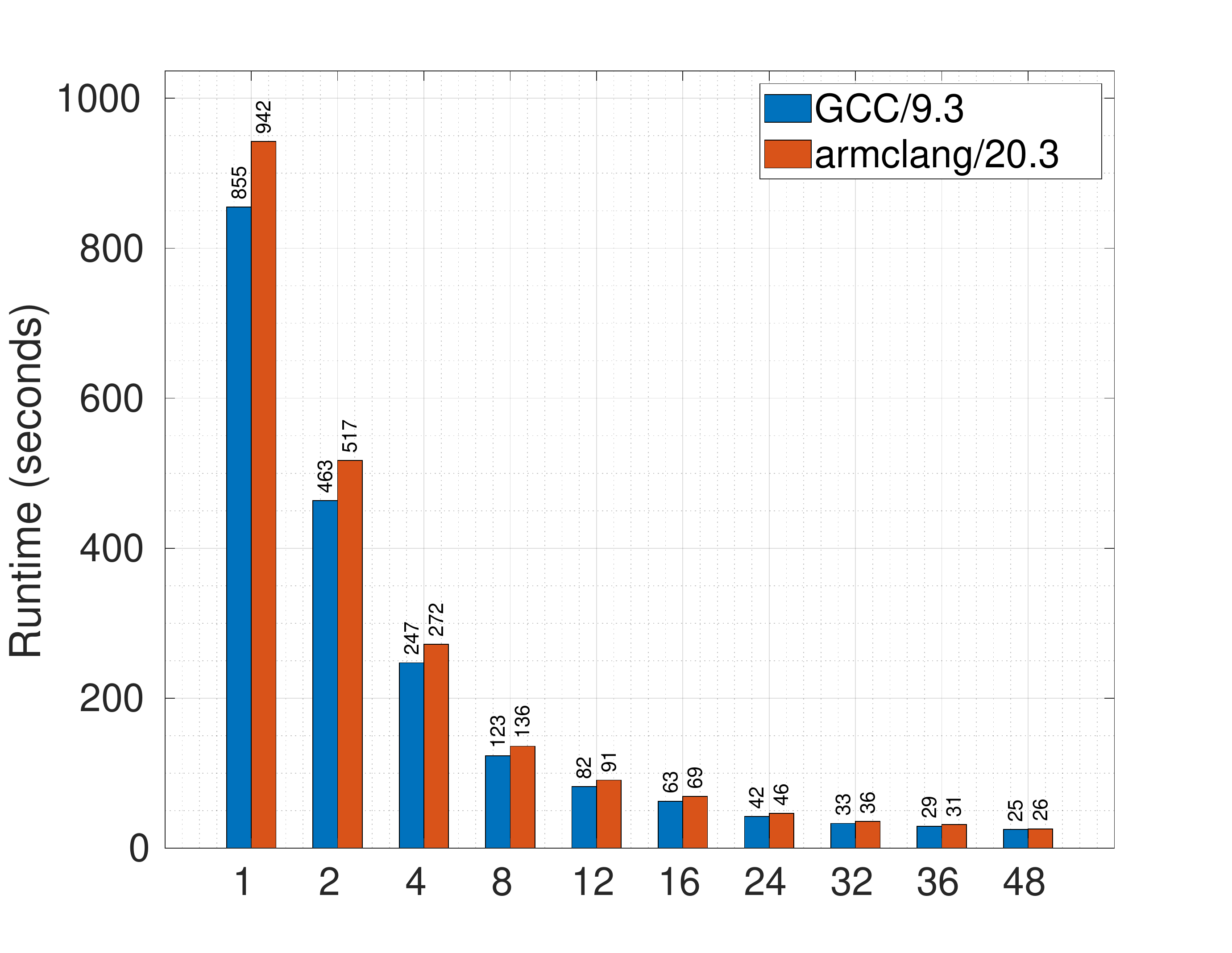}\label{fig:minimod-loopxy}}
    \subfigure[]{\includegraphics[width=0.411\textwidth]{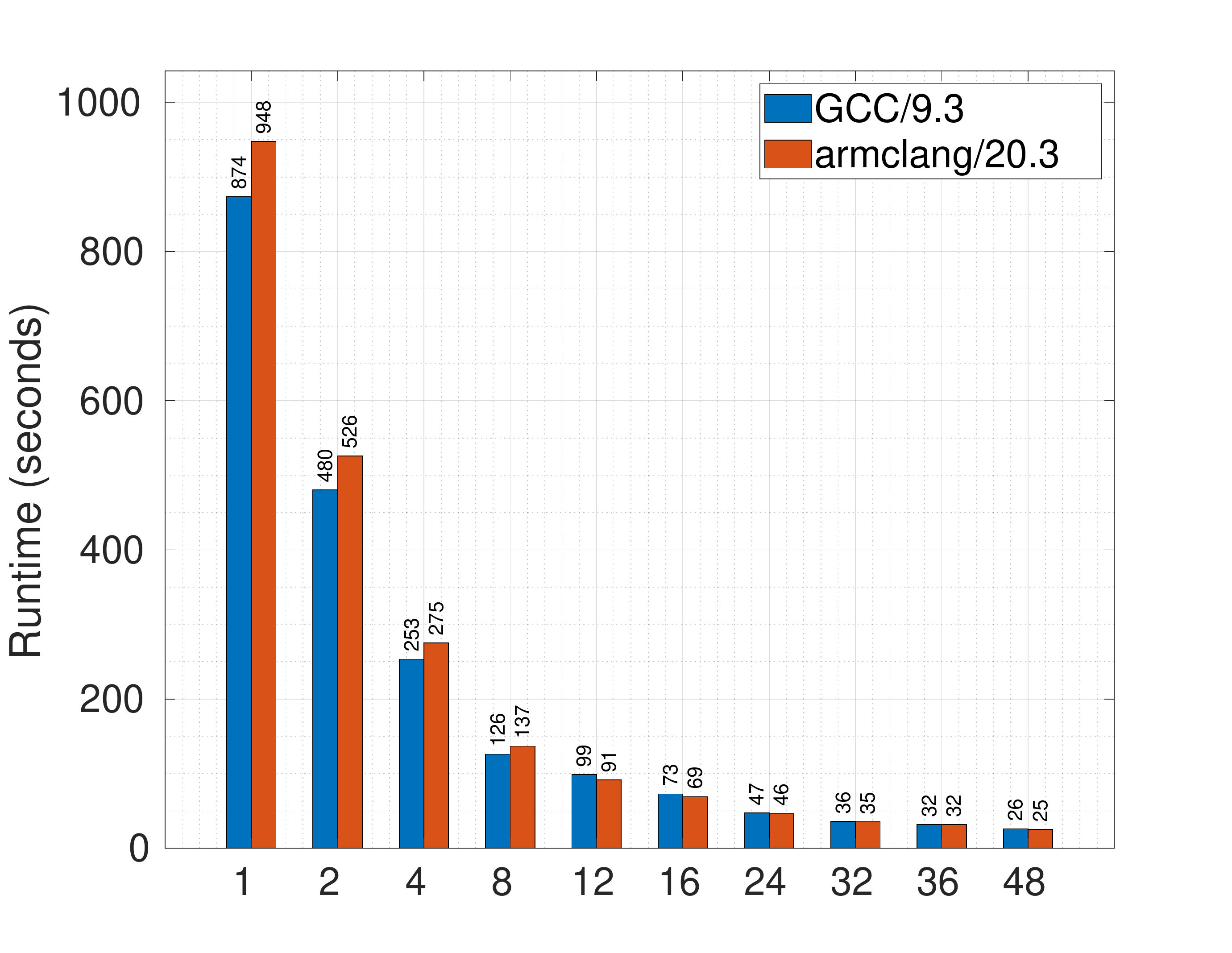} \label{fig:minimod-tasksxy}}
    \caption{(a) Minimod/Ookami/timing (loop xy) (b) Minimod/Ookami/timing (tasks xy)}
\end{figure}

\begin{figure}[!htb]

  \centering
    \subfigure[]{\includegraphics[width=0.411\textwidth]{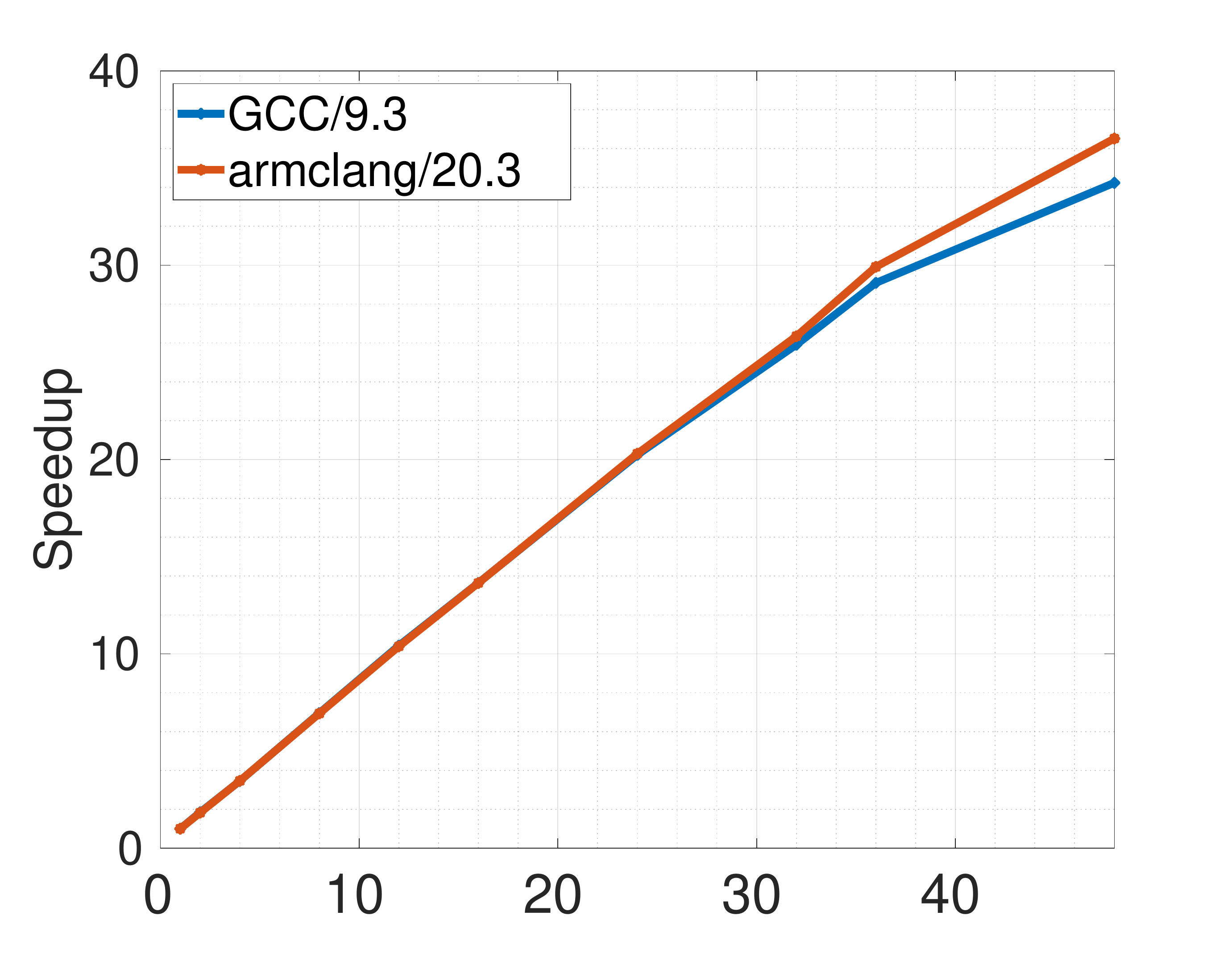}\label{fig:minimod-speedup-loopxy}}
    \subfigure[]{\includegraphics[width=0.411\textwidth]{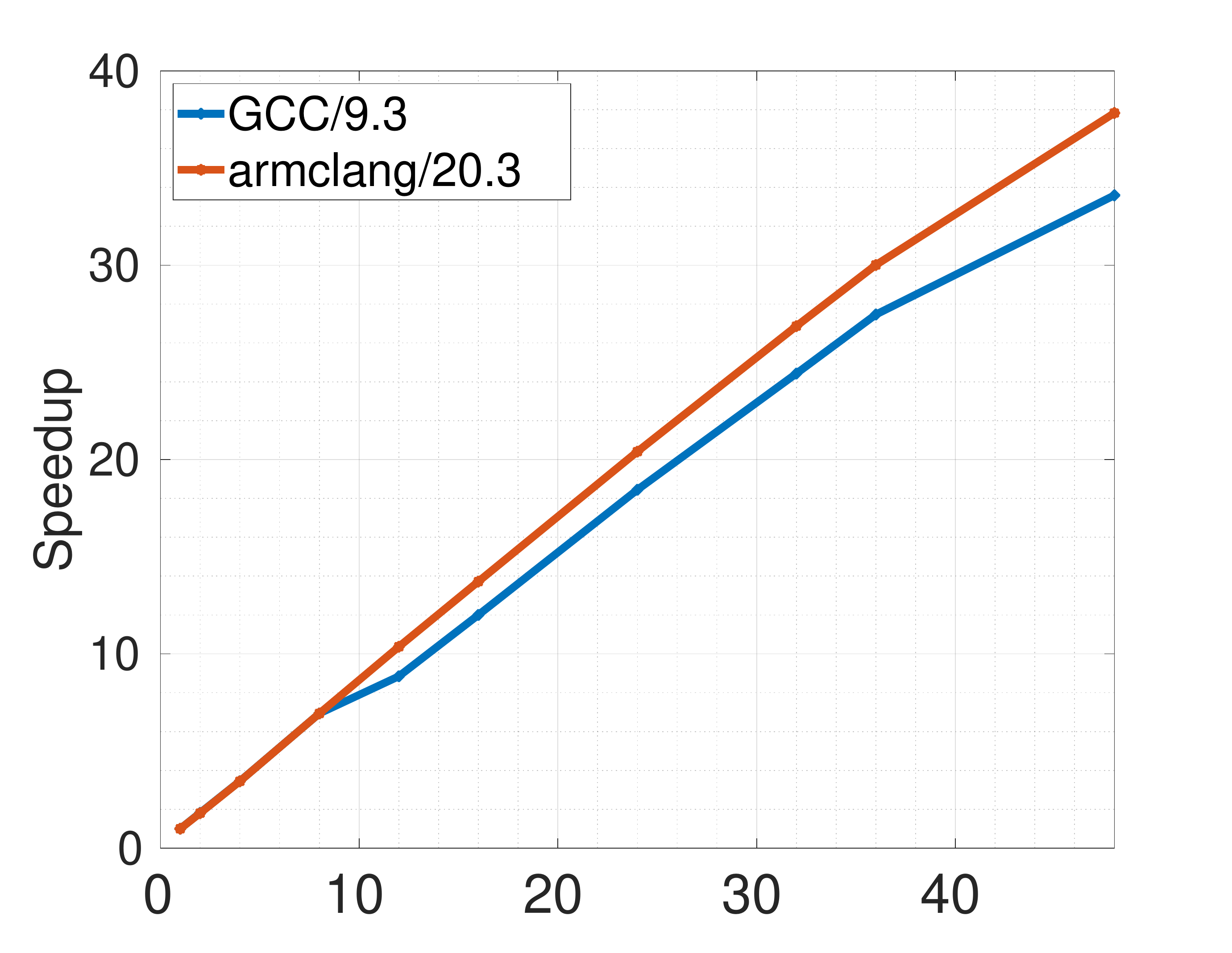} \label{fig:minimod-speedup-tasksxy}}
    \caption{(a) Minimod/Ookami/speedup (loop xy) (b) Minimod/Ookami/speedup (tasks xy)}

\end{figure}

A grid size of $512^3$ was used. Minimod times are shown for each configuration in Figures \ref{fig:minimod-loopxy} and \ref{fig:minimod-tasksxy}, with speedups in \ref{fig:minimod-speedup-loopxy} and \ref{fig:minimod-speedup-tasksxy}. Note that the Cray C compiler was unable to compile this code, due to an internal compiler error, so only GCC and Arm compiler results are shown.\footnote{A Fortran version of Minimod was also evaluated using the Cray Fortran compiler. While this version was successfully compiled, the final numerical result was incorrect with optimization turned on.}

In the GCC compilers, the loop-based configuration tends to outperform the task-based configuration. In LLVM compilers, however, the performance is similar between the two configurations.

We profiled Minimod using the ARM Forge Performance Report tool with 48 threads. We find that in both configurations, the application spends almost the entire runtime within OpenMP regions, and both have a high number of stalled cycles (76.5\% and 80.7\% of cycles for loop-xy and tasks-xy configurations respectively), indicating that the application is memory-bound. This makes the HBM2 memory of the A64FX processor potentially advantageous for this type of application.

%%% Ookami test results here: GFLOPS, Timing, etc.

\subsection{Fugaku}\label{FugakuRes}
With Fugaku's customized Linux kernel and its compute node's processors having 2 extra cores compared to Ookami, it was a slight challenge creating experiments whose environment matched the conditions set in the Ookami-based experiments. The Fujitsu compiler's ability to compile with either their traditional backend and an LLVM backend creates the ability to compare a compiler's performance with itself. In this section, we will break down and explain our results on the Fugaku supercomputer comparing results between GNU Compilers, and the Fujitsu compilers. Per the experiments in Section~\ref{sect:list_of_applications_and_setup}, we ran each application 5 times and took the average of their runtimes. 

\subsubsection{PENNANT}\label{PennantFugaku}
Of all the compilers mentioned in this subsection and in Section~\ref{OokamiRes}, those from Fujitsu resulted in the longest recorded runtimes for both the \textbf{LeblancBig} and \textbf{SedovBig} inputs. In particular, the single-threaded runtimes for both inputs had surprisingly large standard deviations (107 seconds as opposed to a fraction of a second). Conversely, both versions of GNU compilers on Fugaku, when applied to PENNANT, still maintained comparable runtimes to those on Ookami. 

\begin{figure}[!htb]
%%Below: Idea for multiple graphs on one page, each one could have its own label
  \centering
    \subfigure[Compiler Runtime Comparisons]
    {\includegraphics[width=0.411\textwidth]{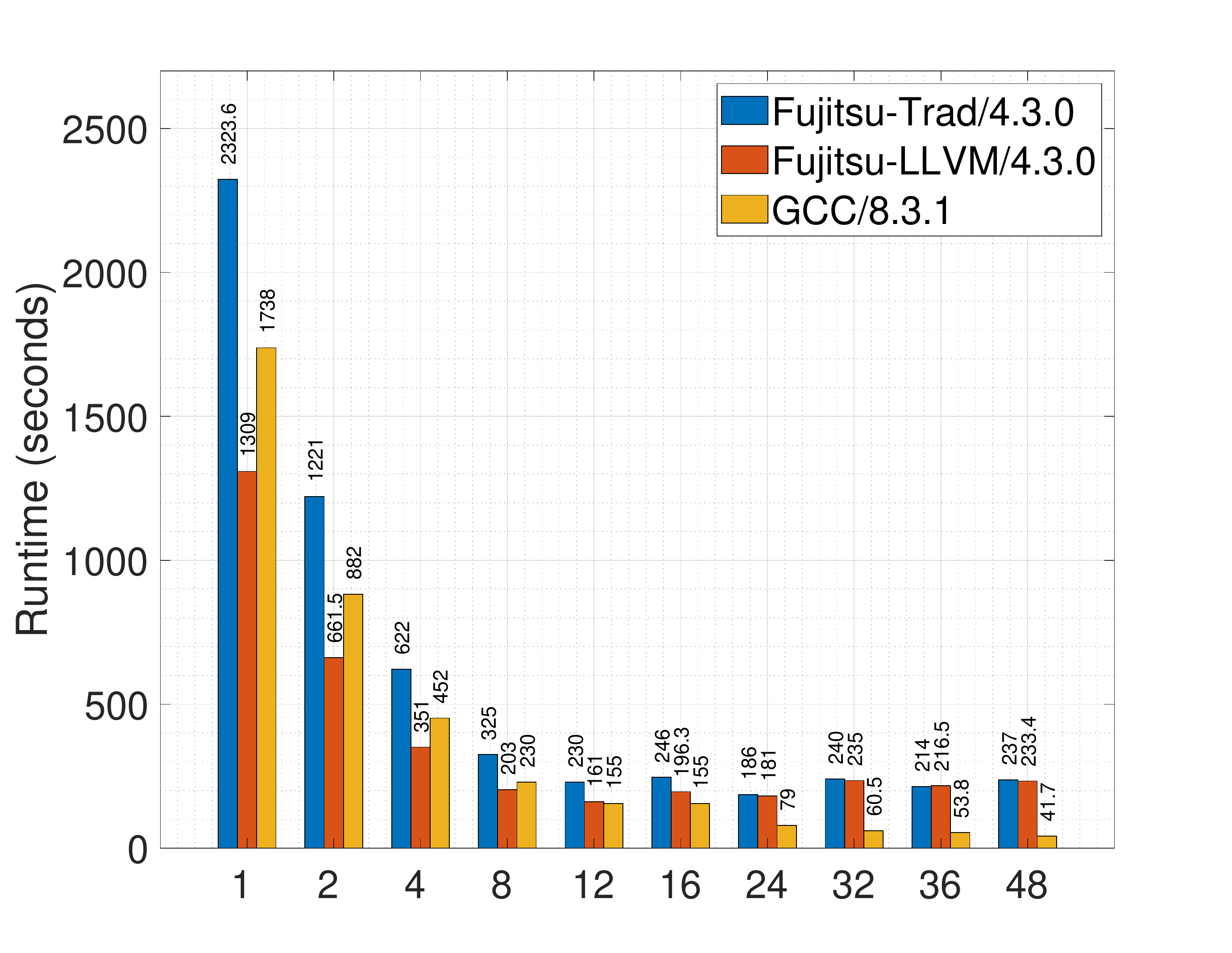} \label{LeblancBig-Fugaku}}
    \subfigure[Relative Speedup]{\includegraphics[width=0.411\textwidth]{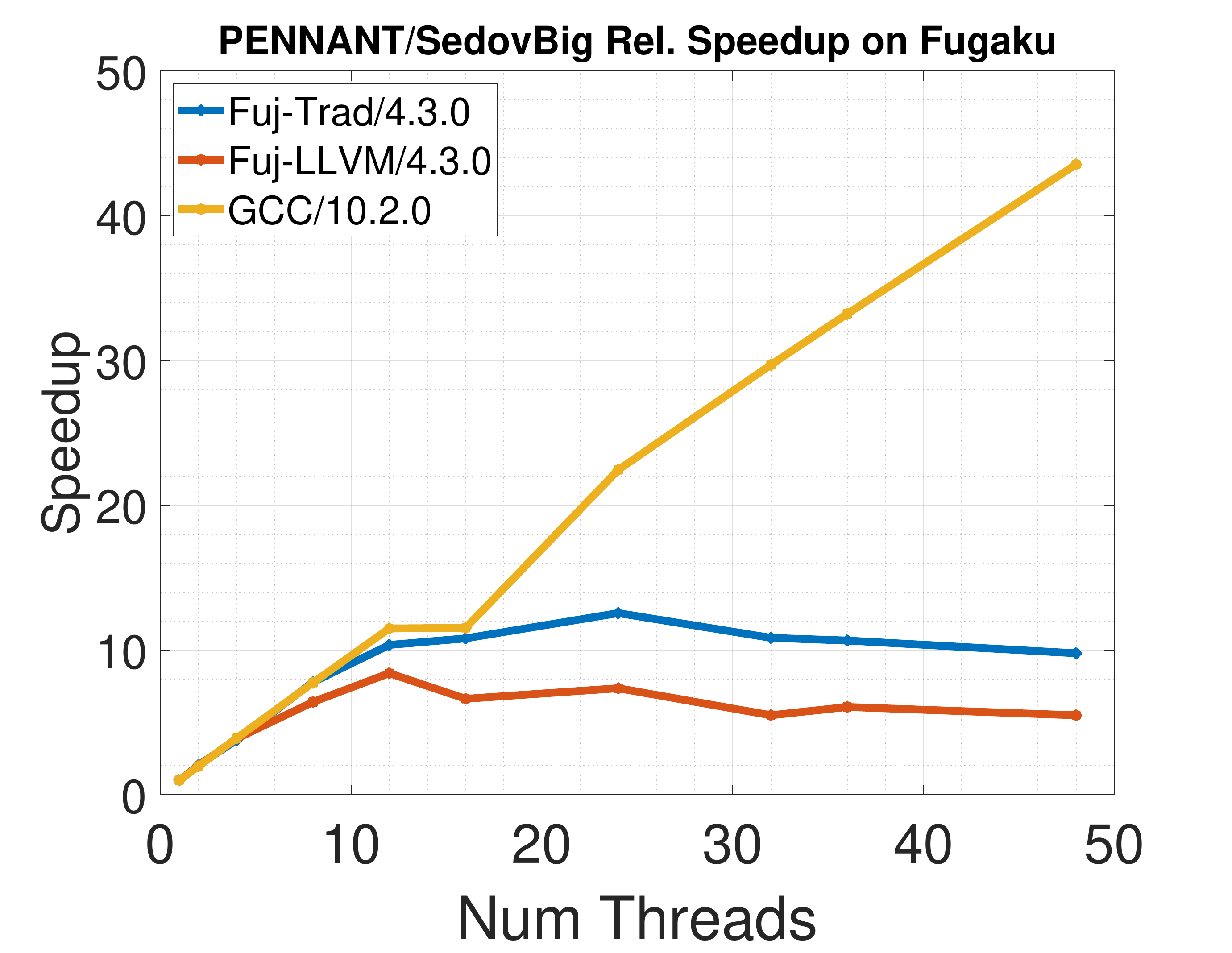}\label{LeblancBig-Fugaku-Speed}}
    
        \subfigure[Compiler Runtime Comparisons]{\includegraphics[width=0.411\textwidth]{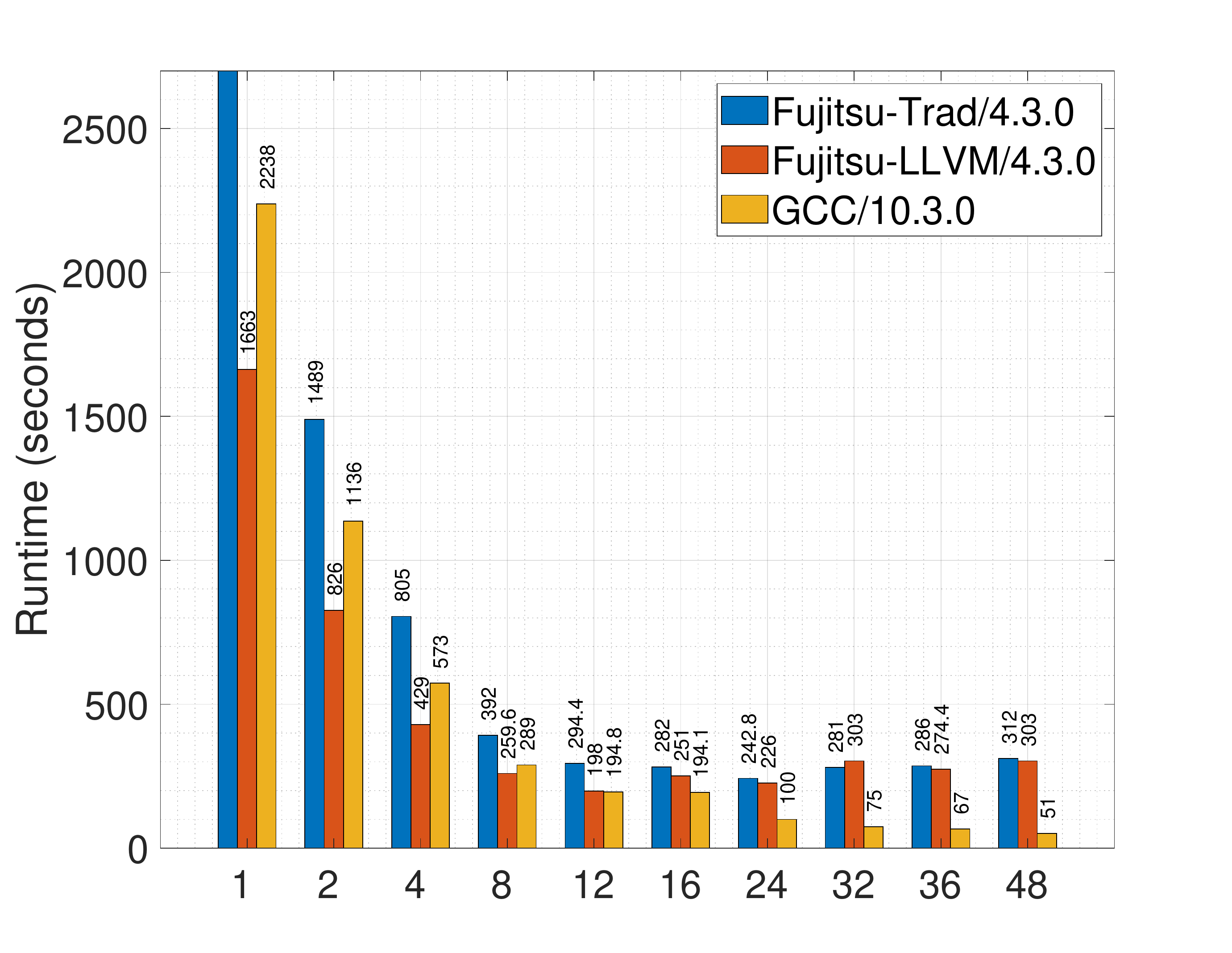} \label{SedovBig-Fugaku}}
    \subfigure[Relative Speedup]{\includegraphics[width=0.411\textwidth]{figures/Pennant-Matlab/Pennant-Fugaku/PENNANT-Speedup-SedovBig-Fugaku.pdf} \label{SedovBig-Fugaku-Speedup}}
    \caption{(a)/(b): LeblancBig Input Results on Fugaku, (c)/(d): SedovBig Input Results on Fugaku; X-axes refer to number of OpenMP threads}
\end{figure}

In Figure~\ref{LeblancBig-Fugaku}, we noticed that the traditional backend options for the Fujitsu Compiler took substantially longer runtimes in 1-and 2-thread runs compared to the LLVM-backend (see Section~\ref{Optimizations}). Profiling \textbf{LeblancBig} shows that the traditional Fujitsu compiler backend takes a longer runtime, yet executes a higher amount of GFLOPS, than the LLVM backend on the \textbf{LeblancBig} input. In particular, both backends show a better runtime at 24 OpenMP threads -- 181 seconds on the LLVM backend and 186 seconds on the traditional backend -- than at 48 OpenMP threads -- 233 and 236 seconds for the LLVM and traditional backends, respectively. This appears to be caused by the increased communication between each of the CMGs on the A64FX processor, especially as the thread count and number of CMGs used increases.  

Similarly, this results in observed reduced speedup for the Fujitsu compiler, especially after reaching 12 threads placed in 1 CMG (See Figure~\ref{LeblancBig-Fugaku-Speed}).

The GNU compilers across both the Ookami and Fugaku systems allow for reasonable speedup, and efficiency as well on top of this. All 3 compiler options first start off with similar trends, but once each compiled binary is run on more than twelve threads, we see a massive drop in efficiency with the Fujitsu compilers. One reason this may be the case is how the underlying communication between CMGs and how the Fujitsu compilers generate SVE instructions to rely more on MPI-based parallelism versus OpenMP/thread-based parallelism, especially if each process only takes 1 thread. 

Similar cases occur with the \textbf{SedovBig} input. Here, we see that 1 CMG full of threads runs more quickly than 2 or more CMGs full of threads with \textbf{LeblancBig}, per in Figure~\ref{SedovBig-Fugaku}. The runtime for a single-OpenMP-thread run with the \textbf{SedovBig} input can take more than 3000 seconds on the Fujitsu-traditional backend, and over half as long as on the Fujitsu-LLVM backend. Similar trends in speedup are shown in Figure~\ref{SedovBig-Fugaku-Speedup}. 

In an experiment on Fugaku, we ran the \enquote{Fujitsu Instant Performance Profiler} (FIPP)\cite{FIPP} on \textbf{SedovBig} using 1 OpenMP thread, as adding more threads results in minimal runtime differences between runs of PENNANT compiled by both Fujitsu backends. Using the LLVM backend results in a shorter runtime (1662 seconds) and fewer GFLOPS (1.41) while having a higher bandwidth usage (1.69 GB/s). Conversely, the traditional backend results in nearly twice the runtime (3108 seconds), a larger GFLOP value (1.93), but just over half the bandwidth usage (0.8868 GB/s) of the LLVM-backend's run. Larger values of OpenMP threads using PENNANT compiled by either backend begins leveling off/converging once a user requests more than 12 OpenMP threads for their application, as per Figure~\ref{SedovBig-Fugaku}.

\subsubsection{SWIM}
\begin{figure}[!htb]
%%Below: Idea for multiple graphs on one page, each one could have its own label
  \centering
    \subfigure[]{\includegraphics[width=0.411\textwidth]{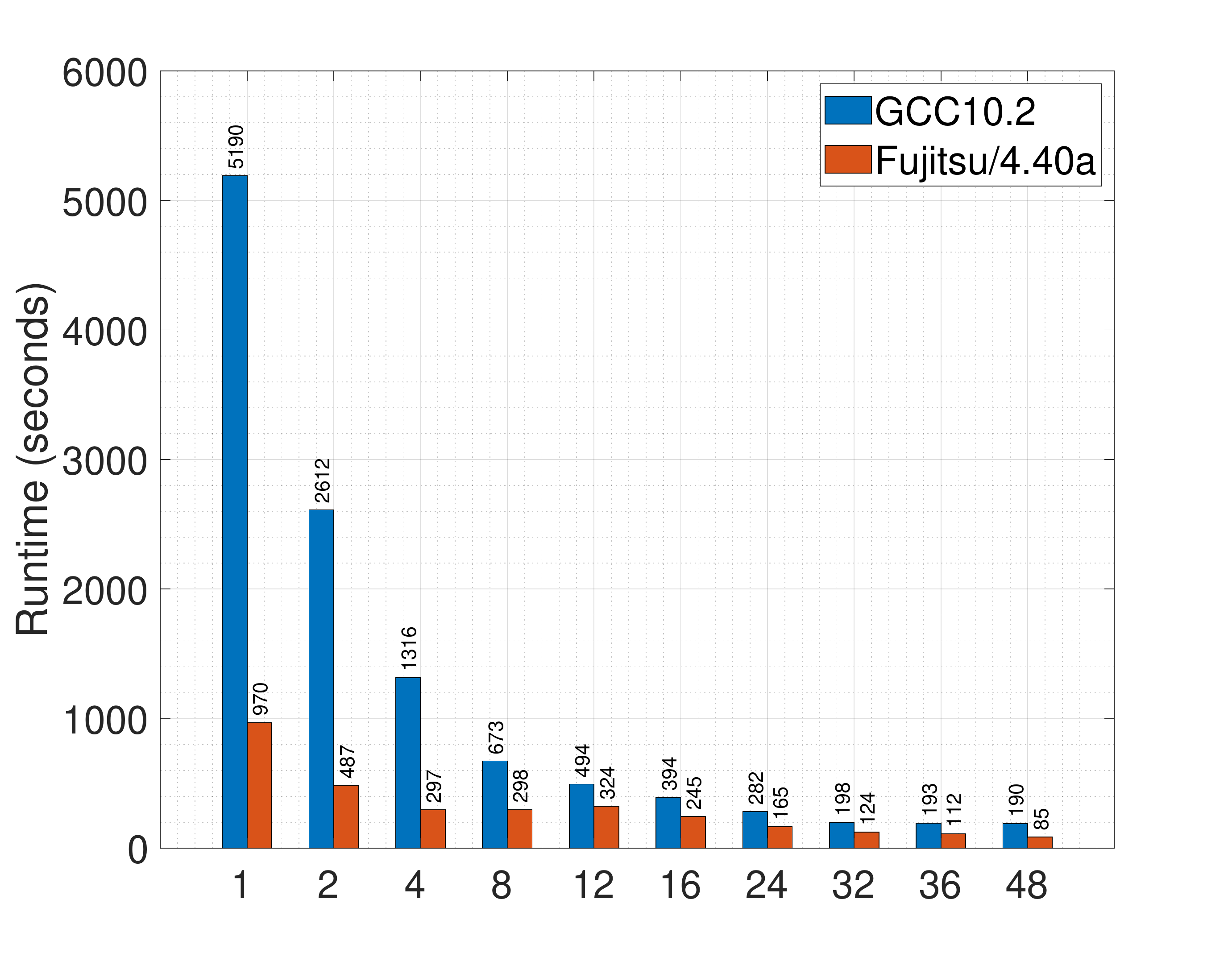} \label{fig:swim-fugaku-runtime}}
    \subfigure[]{\includegraphics[width=0.411\textwidth]{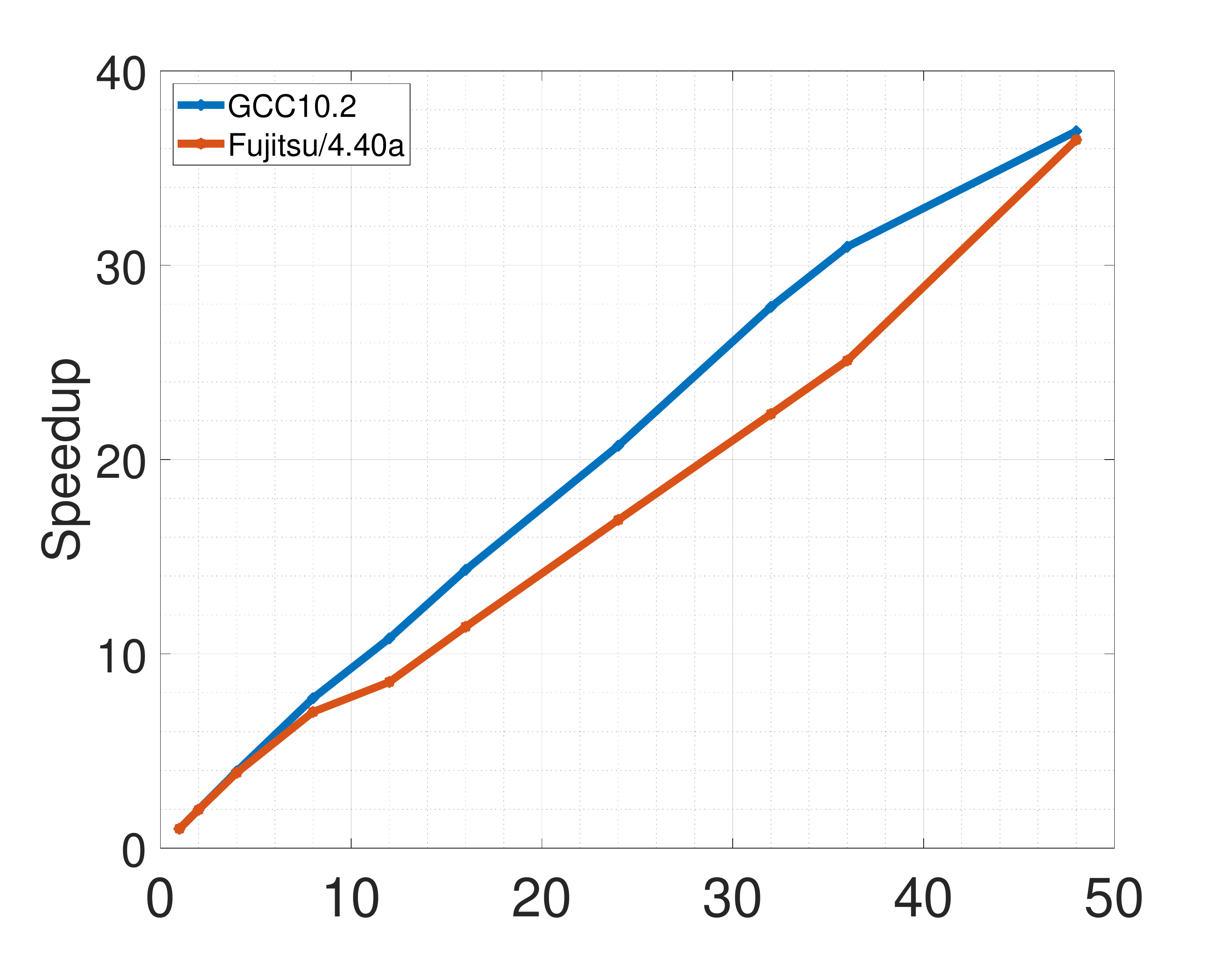} \label{fig:swim-fugaku-Speedup}}
    \caption{(a) SWIM/Fugaku Timing (b) SWIM/Fugaku Relative Speedup}

\end{figure}

With the Fugaku cluster, two compilers were used to test SWIM's capabilities: Fujitsu v4.4.0a and GNU-10.2.1. In contradiction with PENNANT results mentioned in Section \ref{PennantFugaku}, SWIM has gained a significant runtime improvement with Fujitsu compiler compared with best runtime performance with Ookami cluster as shown in Figure ~\ref{fig:swim-fugaku-runtime}. As for the GNU compiler, the runtime results are much comparable with runs made on Ookami. Note that the Fujitsu compilers do not support multiple backends for Fortran.

It is also worth mentioning that memory allocation across multiple CMGs running for thread parallelism is crucial for optimizing SWIM's runtime performance. The environment variable \texttt{\seqsplit{XOS\_MMM\_L\_PAGING\_POLICY}}
is set to \texttt{\seqsplit{demand:demand:demand}} for multiple CMGs in order to place data near the thread that has first touched it, and \texttt{\seqsplit{prepage:demand:prepage}} for a single CMG, as recommended. 

Although the Fujitsu compiler has better overall performance, as shown in Figure~\ref{fig:swim-fugaku-Speedup}, the GNU compiler seems to have a greater speed up than the Fujitsu compiler. GNU runs with 36 threads achieve a 32x speed-up over 1 thread, while he Fujitsu compiler only obtained 25x speed up with the same thread difference. The Fugaku-based runs show a drop in relative speedup starting at 8 OpenMP threads before leveling out at 12 threads.
Our studies on relative compiler efficiency have further backed up our results in  Figure~\ref{fig:swim-fugaku-Speedup}.
% It becomes more clear that Fujitsu compiler tends to better optimize process-based parallelism than thread-based parallelism, which explains why the efficiency of Fujitsu compiler decreases dramatically after 12 threads. 

On Fugaku, with the Fujitsu Instant Performance Profiler, SWIM has shown a much better performance than all other compilers that we have tested previously on Ookami. It has achieved 31.20 GFLOPS with 48 threads, and overall faster runtimes compared to those made on Ookami. One reason might be the extreme high SVE operation rate. As shown in the profiling results, a 99.9\% SVE operation rate has been obtained by Fujitsu compiler. Besides the impressive performance results, the rest are similar with Ookami profiling results. Most of the runtime went into the OpenMP region.
%and high threads running might lead into runtime overheads. 

\subsubsection{Minimod} 
\begin{figure}[!htb]
%%Below: Idea for multiple graphs on one page, each one could have its own label
  \centering
    \subfigure[]{\includegraphics[width=0.411\textwidth]{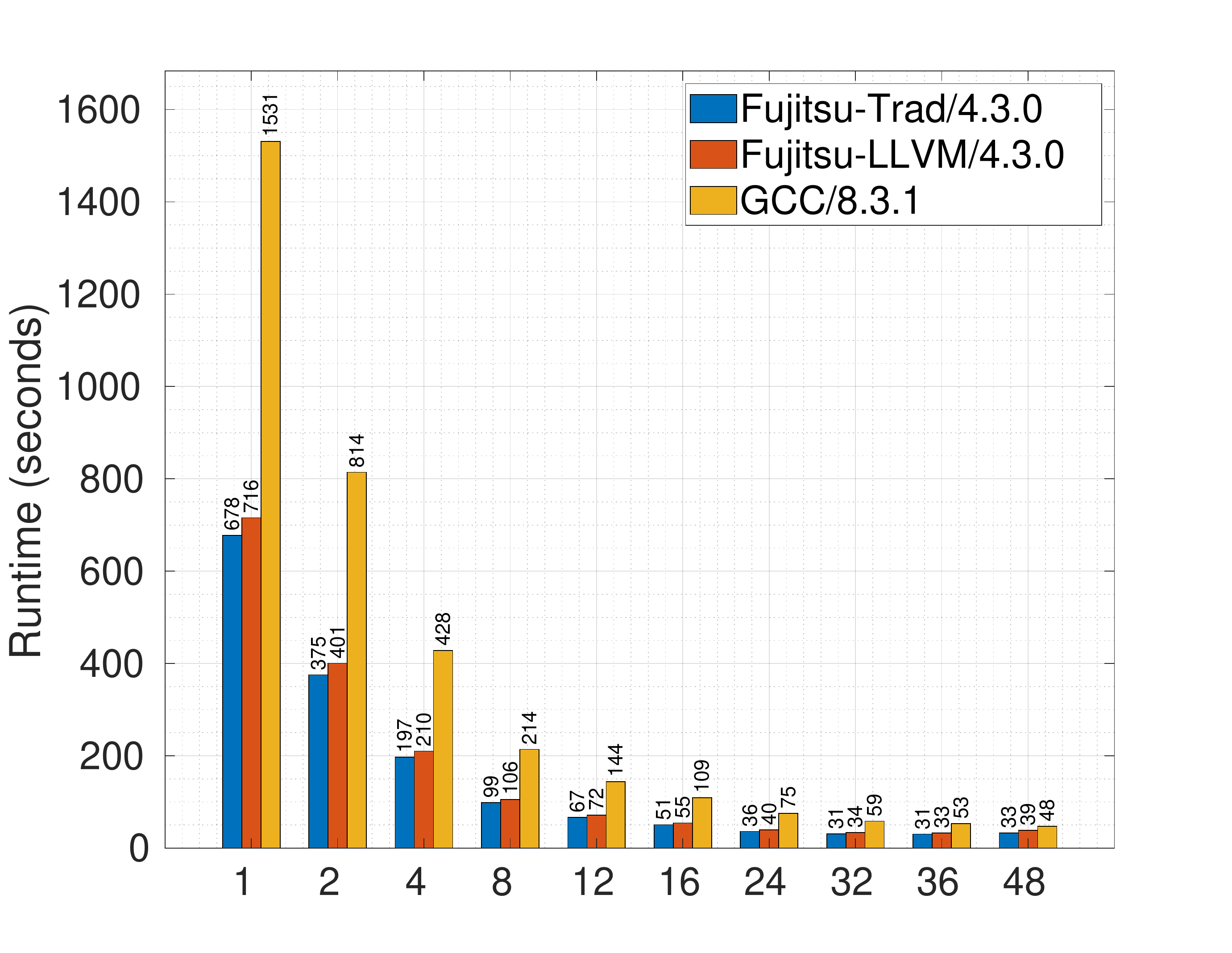} \label{fig:minimod-loopxy-fugaku}}
    \subfigure[]{\includegraphics[width=0.411\textwidth]{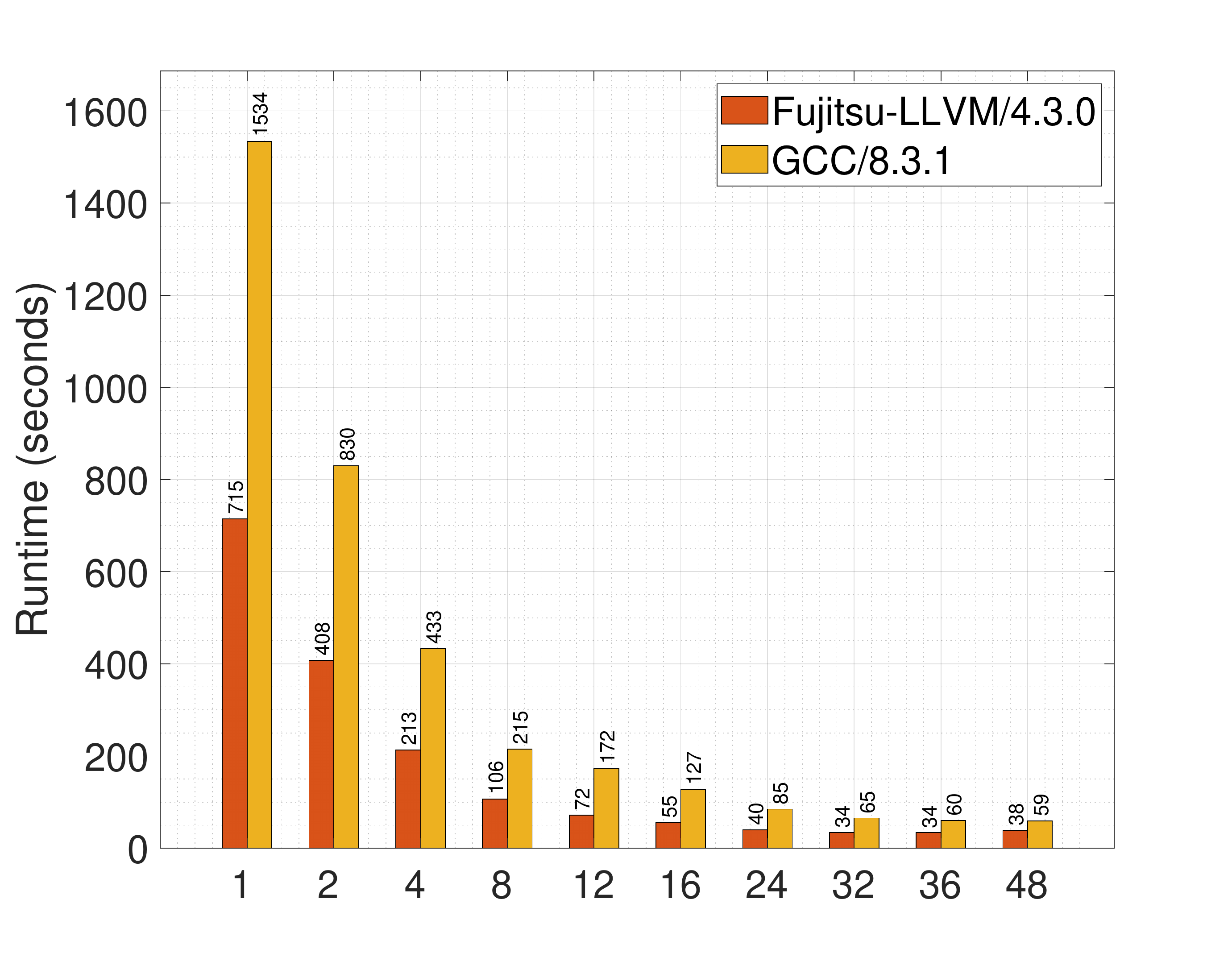} \label{fig:minimod-tasksxy-fugaku}}
    \caption{(a) Minimod/Fugaku/timing (loop xy) (b) Minimod/Fugaku/timing (tasks xy)}

\end{figure}

\begin{figure}[!htb]
%%Below: Idea for multiple graphs on one page, each one could have its own label
  \centering
    \subfigure[]{\includegraphics[width=0.411\textwidth]{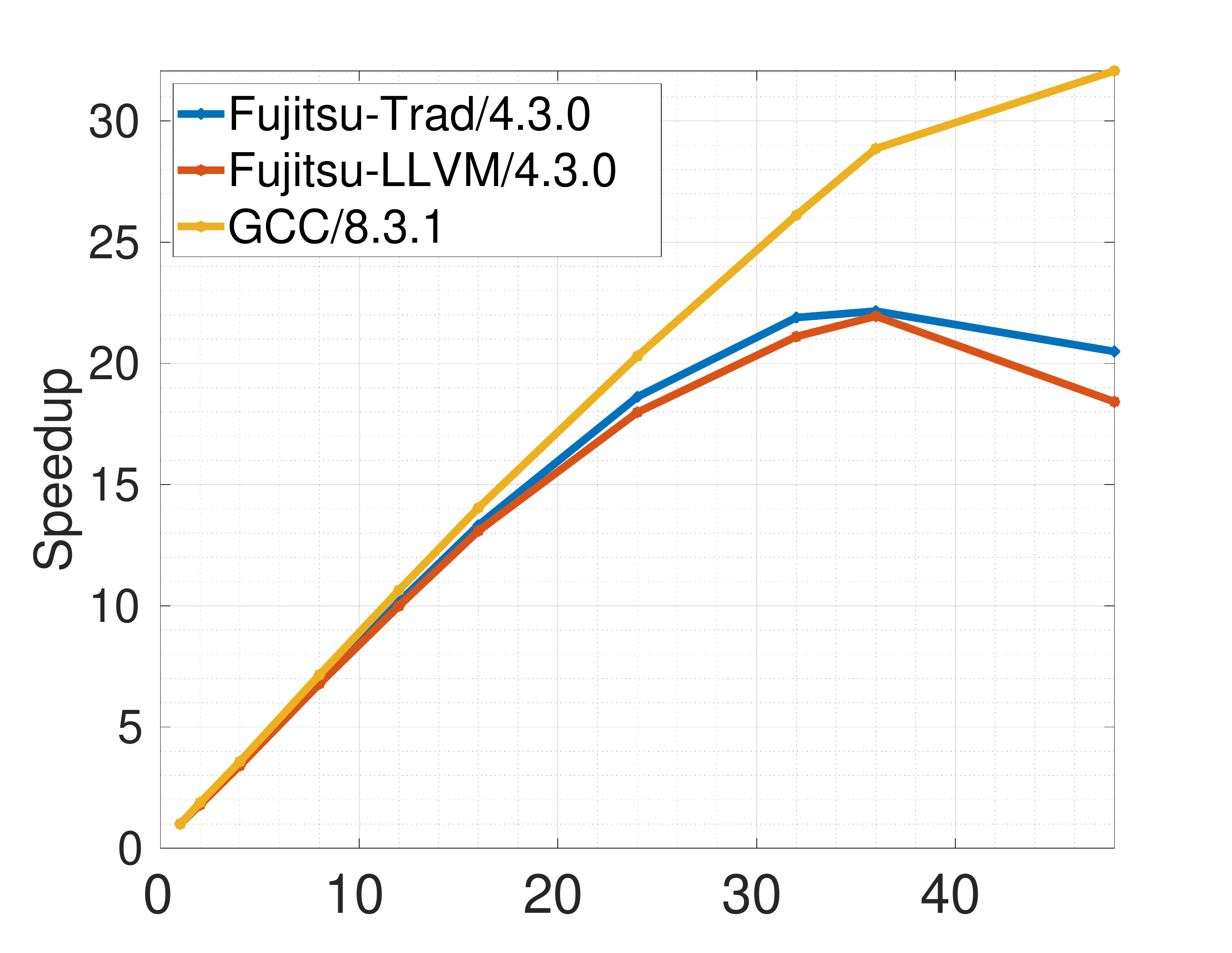} \label{fig:minimod-speedup-loopxy-fugaku}}
    \subfigure[]{\includegraphics[width=0.411\textwidth]{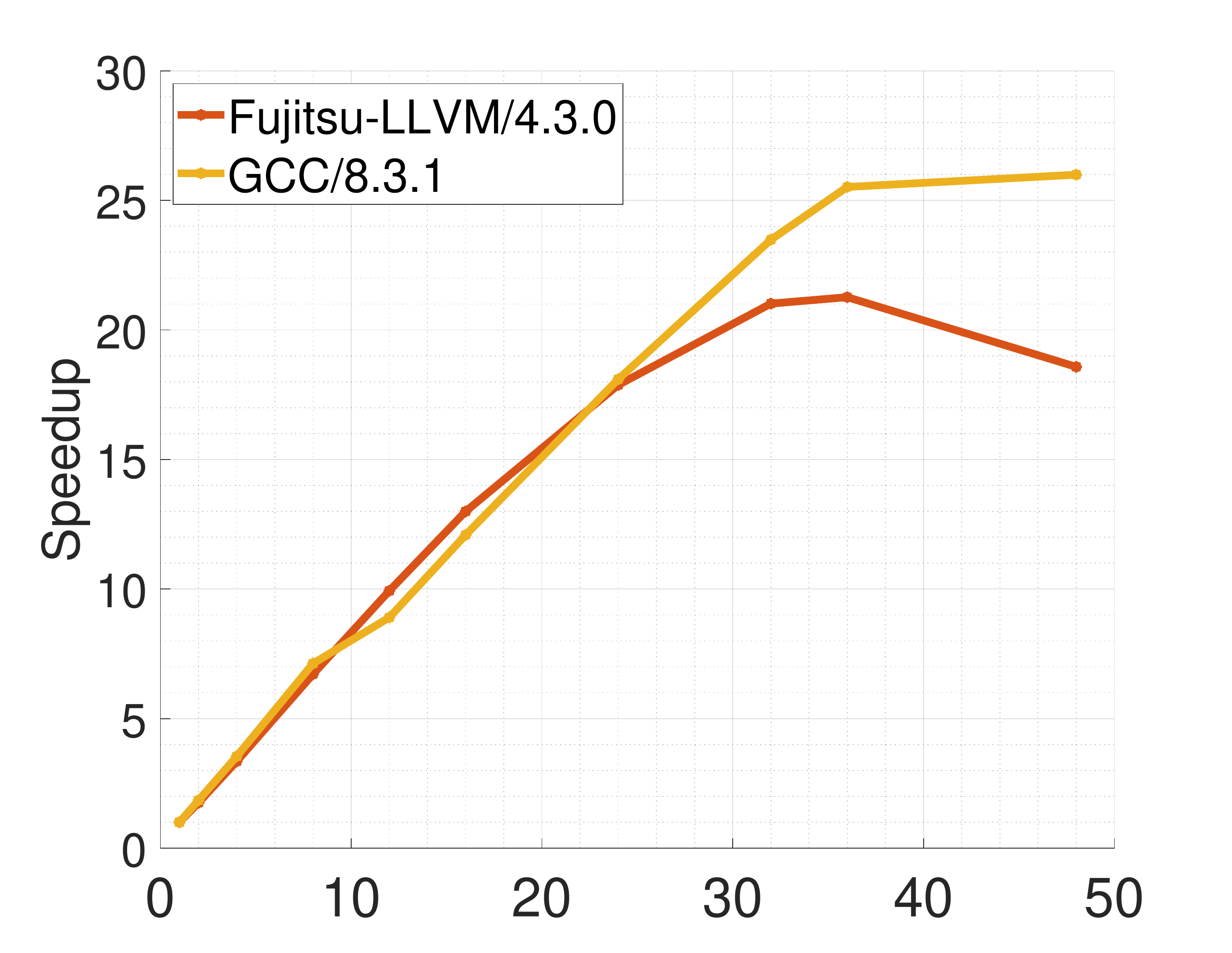} \label{fig:minimod-speedup-tasksxy-fugaku}}
    \caption{(a) Minimod/Fugaku/speedup (loop xy) (b) Minimod/Fugaku/speedup (tasks xy)}

\end{figure}

times are shown for each configuration in Figures \ref{fig:minimod-loopxy-fugaku} (loop xy) and \ref{fig:minimod-tasksxy-fugaku} (tasks xy), with speedups in \ref{fig:minimod-speedup-loopxy-fugaku} and \ref{fig:minimod-speedup-tasksxy-fugaku}. Because the traditional backend for the Fujitsu compiler supports OpenMP up to only version 3.0, it cannot compile the task-based version of Minimod, whereas the LLVM backend supports up through the latest OpenMP specification versions, per Figures \ref{fig:minimod-tasksxy-fugaku} and \ref{fig:minimod-speedup-tasksxy-fugaku}.

Profiling of the Minimod application on Fugaku is currently in progress.

\section{Related Work} \label{sect:related_work}
In~\cite{Odajima2020}, a group from RIKEN reports their preliminary performance analysis of A64FX compared to the Marvell (Cavium) ThunderX2 (TX2) and Intel Xeon Skylake (SKL) processors based on 7 HPC applications and benchmarks. Some of the applications considered use only OpenMP and others use hybrid MPI + OpenMP for parallelization. The compilers used in this study are the Fujitsu Compiler 4.2.0 (under development) for A64FX, ARM-HPC Compiler 20.1 for TX2, and Intel Compiler 19.0.5.281 for SKL.

Another group~\cite{Jackson2020} from EPCC at The University of Edinburgh, reports on their study of various complex scientific applications and mini-kernel benchmarks across multiple nodes, as well as on a single node on different production HPC platforms, which include Fujitsu A64FX processors, 3 Intel Xeon series -- E5-2697 v2 (IvyBridge), E5-2695 (Broadwell), and Platinum 8260M (Cascade Lake)-- and Marvell ThunderX2. Different compiler families, including several versions of some of them, like, Fujitsu, Intel, GCC/GNU, ARM/LLVM, and Cray, were used in their study. Also, they have considered various MPI implementations and
scientific libraries.  

Several recent works evaluated benchmark applications using multiple compilers on the A64FX processor; e.g., \cite{poenaru2021evaluation,graziano2021optimizing}. However, these works do not focus on OpenMP.

\section{Conclusions and Future Work} \label{sect:conclusions_and_future_work}
In this paper, we have studied and observed the behavior of OpenMP
implementations
on the A64FX processor across several applications and several compiler toolchains. We have observed that Cray's compilers and GNU/LLVM compilers that have support for ARM-based processors appear to scale better with OpenMP compared to the Fujitsu compilers. We have observed that, while having the most optimal performance, the Cray Compilers may fail to compile code or generate incorrect instructions, such as with the Minimod application, leading to incorrect results. 

Moving forward, we wish to explore more complex OpenMP behavior, including different data-sharing attributes and SIMD clauses. In addition, examining how the Fujitsu compiler toolchain behaves on Ookami will make an interesting comparison between both its structure and that on Fugaku. Another avenue to explore would be to examine serial versus OpenMP runtimes and analyze how much of an impact the overhead has in each runtime environment across compilers.

\section{Acknowledgements}
We would like to thank the NSF for supporting the Ookami cluster, and the ability to research the A64FX processor by Riken and Fujitsu, through grant OAC 1927880. We would like to thank the Riken Center for Computational Science for providing us with accounts to use the Fugaku supercomputer and conduct research on it. We would also like to thank Stony Brook University and the Institute for Advanced Computational Science for providing the resources to allow us to conduct our studies on Ookami. Finally, we would like to thank TotalEnergies Exploration and Production Research and Technologies for their support of experimentation using MiniMod.

\clearpage
\bibliographystyle{abbrv}
\bibliography{references}

%\vspace{12pt}
%\color{red}
%IEEE conference templates contain guidance text for composing and formatting conference papers. Please ensure that all template text is removed from your conference paper prior to submission to the conference. Failure to remove the template text from your paper may result in your paper not being published.

\end{document}